# Requirements for functional *pn*-homojunctions in lead-halide perovskite solar cells

Thomas Kirchartz[1,2*], David Cahen[3]

[1]IEK5-Photovoltaik, Forschungszentrum Jülich, 52425 Jülich, Germany
[2]Fac. of Engineering and CENIDE, Univ. of Duisburg-Essen, Carl-Benz-Str. 199, 47057 Duisburg, Germany
[3]Department of Materials and Interfaces, Weizmann Institute of Science, Rehovoth 76100, Israel

* author for correspondence, email: t.kirchartz@fz-juelich.de

Cui et al.[1] describe the fabrication and characterization of planar *pn*-junction solar cells based on lead-halide perovskites. The doping densities measured using Hall effect measurements vary from $N_D = 1\times10^{12}$ cm$^{-3}$ to $8\times10^{12}$ cm$^{-3}$ for the solution-processed *n*-type layer and $N_A = 8\times10^{9}$ cm$^{-3}$ for the evaporated *p*-type layer. While these devices outperform their counterparts, that are supposedly un-doped, the results raise three important questions: (i) Are the reported doping densities high enough to change the electrostatic potential distribution in the device from that for the un-doped ones, (ii) are the doping densities high enough for the *pn*-junction to remain intact under typical photovoltaic operation conditions and (iii) is a *pn*-junction beneficial for photovoltaic performance given the typical properties of lead-halide perovskites.

The first two questions can be answered quite easily by considering basic semiconductor device physics. The first criterion that the doping densities of the two layers of a *pn*-junction have to satisfy is to ensure that the depletion width $w = \sqrt{2\varepsilon_0\varepsilon_r V_{bi}/qN_{A/D}}$ caused by the doping is substantially smaller than the total absorber layer thickness *d*. Thus, the doping densities have to fulfill $N_{A/D} > 2\varepsilon_0\varepsilon_r V_{bi}/qd^2$, which yields a minimum doping concentration of ~$10^{16}$ cm$^{-3}$ when using typical values for the built-in voltage $V_{bi} \approx 1$ V, a relative permittivity $\varepsilon_r \approx 30$ [2] and the reported thickness $d \approx 500$ nm. Note that while we do not know the precise value of $V_{bi}$ for the present or other perovskite solar cell geometries, it cannot be much smaller than the voltage at the maximum power point in order to avoid S-shaped *I-V* curves[3] and, therefore, has to be around 1 V or higher. Even using the highest doping density mentioned in ref. [1], we obtain depletion widths of ~20 µm (c.f. Fig. S4), i.e. much wider than any halide perovskite thin film in these solar cells and also wider than basically any thin-film solar cell made of any material considered so far in photovoltaics.

This lower limit of ~$10^{16}$ cm$^{-3}$ for the doping concentration applies at thermal equilibrium, e.g. 0 V in the dark. At typical operating conditions of a solar cell, photogenerated charge carriers are created and if their concentrations, *n* and *p*, exceed the doping concentrations, the *pn*-junction will disappear. At the maximum power point, the product *np* of electron and hole densities is $np \geq n_i^2 \exp(qV_{mpp}/kT)$, where $n_i$ is the intrinsic carrier concentration, $V_{mpp}$ is the voltage at the maximum power point and $kT/q$ is the thermal voltage. The ≥ sign is necessary because the quasi-Fermi level splitting inside the device is always slightly higher than the externally measured voltage due to finite quasi-Fermi level gradients, needed to drive current.[4] For the doping densities to be still effective at $V_{mpp}$, the product $N_D N_A$ has to exceed $np$ at the maximum power point. The $V_{mpp}$ in ref. [1] is 0.89 V leading to $np \approx 6.1 \times 10^{24}$ cm$^{-6}$ which is about two orders of magnitude higher than the product $N_D N_A \approx 6.4 \times 10^{22}$ cm$^{-6}$. So while the measured doping density is still too low to exceed our estimate of $np$, this criterion is indeed less selective than the first one. Thus, doping densities ~$10^{16}$cm$^{-3}$ of donors and acceptors would have been able to satisfy both criteria.

To sum up our observations, Figure 1 shows the band diagrams under one sun illumination at $V$ = 0.89 V, simulated with the software SCAPS,[5,6] for

**a** the doping density stated in ref. [1],

**b** $10^{16}$ cm$^{-3}$ and

**c** $10^{17}$ cm$^{-3}$ (equal for donor and acceptor concentrations in the respective layers). The band diagram in panel **a** corresponds exactly to the one without doping (see Figs. S1 and S2). Due to the band alignment assumed in the simulation, there is an excess of electrons leading to a convex shape of the conduction and valence band edge. In panels **b** and **c** the doping becomes apparent with the conduction band edge being visibly closer to the electron quasi-Fermi level in the entire *n*-type region (blue shaded) than in the *p*-type region (red shaded region).

Another way of verifying that the obtained doping densities are insufficient to achieve a functional *pn*-junction at relevant photovoltaic operating conditions is to compare the difference in Fermi levels determined using X-ray photoemission spectroscopy and shown in Fig. 3b of ref. [1] for films that are made to be similar to the *n*- and *p*-type regions. The maximum difference in Fermi levels shown is 0.73 eV, which is fairly consistent with the observed doping densities, but is again substantially less than $qV_{mpp}$ = 0.89 eV.

While the process used to obtain the purported *pn*-junction in ref. [1] undeniably leads to higher efficiencies, we can be sure that it is not the doping that causes this improvement. In addition, we note that the drift-diffusion simulation shown in Fig. 2c is performed using doping densities > $10^{17}$ cm$^{-3}$ (see supplementary table 4 in ref. [1]); indeed, such densities would suffice to create a real *pn* junction (cf. our Fig. 1 **c**), but they are orders of magnitude higher than the measured values reported in ref. [1].

A challenging question is the third one posed above, namely whether a *pn*-junction in lead-halide perovskite solar cells is a target worth pursuing. This question has no generic answer, because the ideal band diagram for a solar cell depends heavily on the material properties. If the diffusion length is long enough and charge collection is efficient as is the case in high efficiency lead-halide perovskite solar cells, the key question for maximizing the voltage per extracted carrier is how high the recombination rate is for a given defect density and capture cross section and how high the ideality factor is. The former would reduce the open-circuit voltage, while ideality factors > 1 would reduce the fill factor.[7,8] Both losses are maximized for the scenario where $n\sigma_n = p\sigma_p$ holds in a large part of the absorber volume,[9] where $\sigma_n$ and $\sigma_p$ are the capture cross sections for electrons and holes into the defect that dominates recombination. In order to avoid the condition $n\sigma_n = p\sigma_p$, doping may be helpful as shown in Fig. 2, because it induces an asymmetry at least in $n$ vs. $p$ that may help to reduce the absorber volume where $n\sigma_n = p\sigma_p$ to a small region.[10] Thus, *pn*-junctions in lead-halide perovskites could indeed be helpful to achieve higher efficiencies if deep defect densities and carrier mobilities will not be adversely affected by higher doping densities and if those higher doping densities can actually be achieved technologically.

This last point may well be key as a *pn*-junction is not a thermodynamically stable situation; rather it is a kinetically stabilized one.[11] The relatively high diffusion coefficients for atomic/ionic species in the Pb halide perovskites that have been reported, will decrease such stabilization; more importantly, and for a defect to persist in a material, any defect formation energy has to be less than the reaction free energy for decomposition of the material,[12] a condition that will limit doping densities beyond those dictated by thermodynamics.

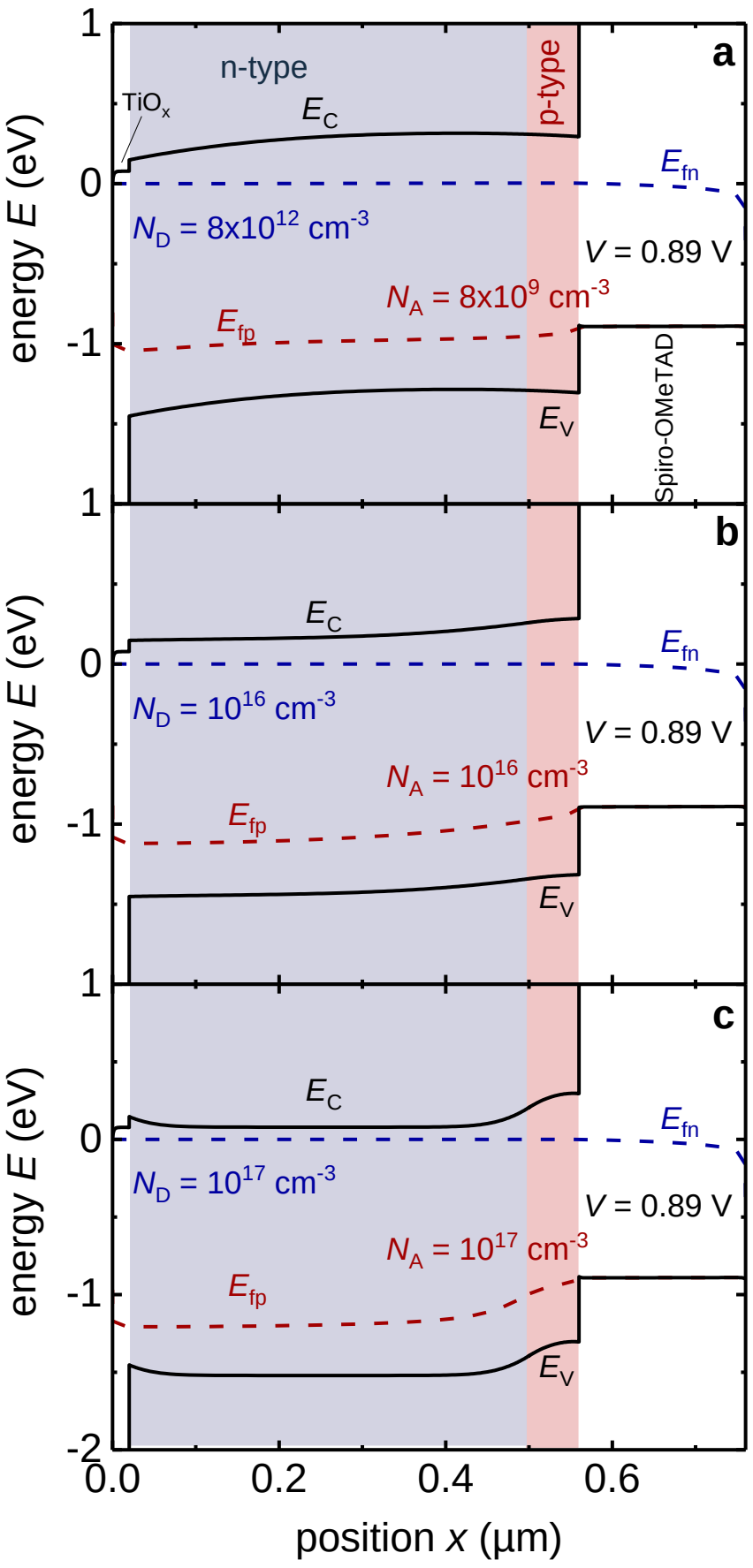

**Fig. 1 | Simulating the effect of doping on the band diagram of halide perovskite solar cells.** Band diagram of a $TiO_x$ (20 nm)/perovskite(*n*) (480 nm)/perovskite(*p*) (60 nm)/SpiroOMeTAD (200 nm) stack with the doping density being **a** as given in ref. [1], **b** $10^{16}$ $cm^{-3}$ for both donor and acceptor concentration and **c** $10^{17}$ $cm^{-3}$ for both donor and acceptor concentration. All other simulation parameters are found in table I in the supplementary information. The simulations are done under 1 sun illumination and at a forward bias of $V$ = 0.89V which corresponds to the maximum power point of the certified cell presented in ref. [1].

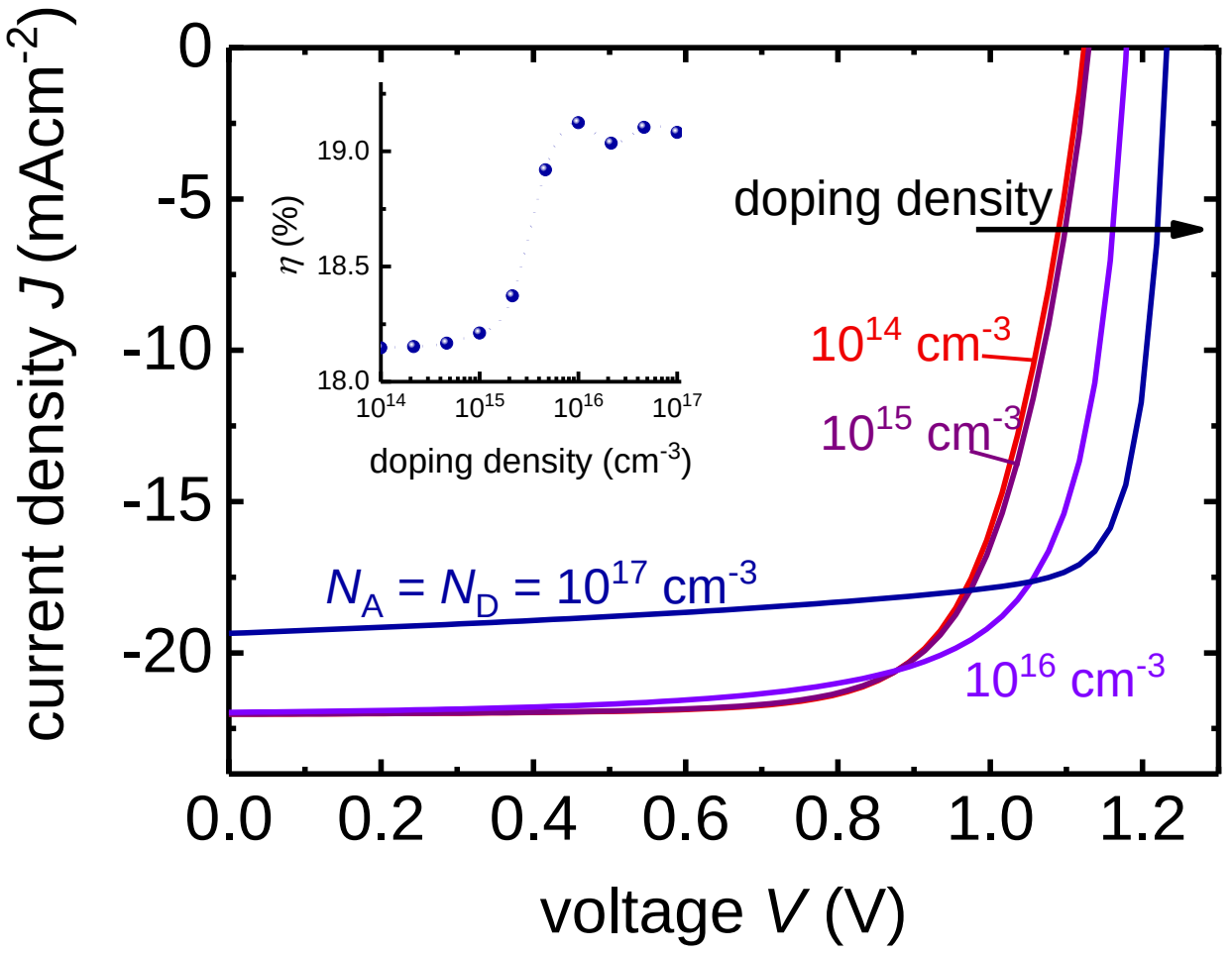

**Fig. 2 | Simulating the effect of doping on perovskite solar cells.** Current-voltage curves, simulated using the same parameters as used for Fig. 1 but with the doping densities varied from $10^{14}$ to $10^{17}$ cm$^{-3}$. Donor and acceptor concentration are always assumed identical. The results show that up to around $10^{15}$ cm$^{-3}$, no effect is visible. For higher doping densities $J_{sc}$ decreases and $V_{oc}$ goes up due to a reduction of the volume where $n\sigma_n=p\sigma_p$. Note that the simulation assumes that in the perovskite, recombination is via a deep trap that has equal capture cross sections for electrons and holes ($\sigma_n=\sigma_p$). Inset: Efficiency as a function of doping density showing a peak at a doping density of $10^{16}$ cm$^{-3}$ that holds for the parameters used here.